\def\year{2021}\relax
\documentclass[letterpaper]{article} 
\usepackage{aaai21}  
\usepackage{times}  
\usepackage{helvet} 
\usepackage{courier}  
\usepackage[hyphens]{url}  
\usepackage{graphicx} 
\def\UrlFont{\rm}  
\usepackage{natbib}  
\usepackage{caption} 
\usepackage{amsmath, amsthm, amsfonts} 
\usepackage{multirow, makecell}
\usepackage{lineno}
\usepackage[linesnumbered,ruled,vlined,english,onelanguage]{algorithm2e}
\usepackage{comment}
\usepackage{xcolor}
\title{Enabling Fast and Universal Audio Adversarial Attack Using Generative Model }
\author{
    Yi Xie$^1$, Zhuohang Li$^2$, Cong Shi$^{1}$, Jian Liu$^2$, Yingying Chen$^1$, Bo Yuan$^{1}$
    \\}
\affiliations{$^1$Rutgers University\\ $^2$The University of Tennessee, Knoxville\\
yi.xie@rutgers.edu, zli96@vols.utk.edu, cs1421@winlab.rutgers.edu,\\ jliu@utk.edu, yingche@scarletmail.rutgers.edu, bo.yuan@soe.rutgers.edu}

\begin{document}

\maketitle

\begin{abstract}

Recently, the vulnerability of deep neural network (DNN)-based audio systems to adversarial attacks has obtained increasing attention. However, the existing audio adversarial attacks allow the adversary to possess the entire user's audio input as well as granting sufficient time budget to generate the adversarial perturbations. These idealized assumptions, however, make the existing audio adversarial attacks mostly impossible to be launched in a timely fashion in practice (e.g., playing unnoticeable adversarial perturbations along with user's streaming input). To overcome these limitations, in this paper we propose fast audio adversarial perturbation generator (FAPG), which uses generative model to generate adversarial perturbations for the audio input in a single forward pass, thereby drastically improving the perturbation generation speed. Built on the top of FAPG, we further propose universal audio adversarial perturbation generator (UAPG), a scheme to craft universal adversarial perturbation that can be imposed on arbitrary benign audio input to cause misclassification. Extensive experiments on DNN-based audio systems show that our proposed FAPG can achieve high success rate with up to $214\times$ speedup over the existing audio adversarial attack methods. Also our proposed UAPG generates universal adversarial perturbations that can achieve much better attack performance than the state-of-the-art solutions.

\end{abstract}

\section{Introduction}
\label{sec:intro}

As the current most powerful artificial intelligence (AI) technique, deep neural networks (DNNs) have been widely adopted in many practical applications. Despite their current success and popularity, DNNs suffer from several severe limitations, especially the inherent high vulnerability to adversarial attack~\cite{goodfellow2014explaining,carlini2017towards}, a very harmful attack approach that imposes well-crafted adversarial perturbation on the benign input of DNNs to cause misclassification. Being originally discovered in the image classification applications, to date the vulnerability of DNNs, especially various types of adversarial perturbation generation methods~\cite{kurakin2016adversarial,poursaeed2018generative,moosavi2017universal}, has been extensively investigated in many image-domain applications.

Considering the rapidly increasing use of DNNs in modern audio-domain applications and systems, such as smart speaker (e.g., Apple Homepod, Amazon Echo) and voice assistant (e.g., Siri, Google Assistant, Alexa), recently both machine learning and cybersecurity communities have begun to study the possibility of adversarial attack in the audio domain. Some pioneering efforts \cite{carlini2018audio,neekhara2019universal} in this topic have demonstrated that the idea of injecting inconspicuous perturbations into benign voice inputs to mislead the DNN-based audio systems is not just conceptually attractive but also practically feasible. To date, several works have reported the successful adversarial attacks in different audio-domain applications, including but not limited to speaker verification \cite{kreuk2018fooling,chen2019real}, speech command recognition \cite{alzantot2018did,DBLP:journals/corr/abs-1905-13399}, speech-to-text transcription \cite{carlini2018audio,yuan2018commandersong}, and environmental sound classification \cite{abdoli2019universal}.

\textbf{Limitations of Prior Work.} Although the existing work have already demonstrated the feasibility of audio adversarial attack, they are still facing several challenges. More specifically, the state-of-the-art audio adversarial attack approaches make several idealized assumptions on the attacking setting: 1) \textit{Having large time budget for generating adversarial perturbation.}
In practical audio applications, the benign inputs are typically quickly-streaming voice input. Therefore, due to such temporal constrain, the existing audio adversarial attacks, which rely on time-consuming iterative optimization approaches such as C\&W~\cite{carlini2018audio} or genetic algorithms~\cite{alzantot2018did}, are too slow to launch the attack against these real-time audio processing systems; 2) \textit{Owning authorization to observe the context of the benign input}. Since the existing perturbation generation methods require to pre-know the full content of the ongoing voice input, the inherent sequential nature of audio signals makes it impossible for the adversary to generate adversarial perturbation during input-streaming phase. Consequently, the current audio adversarial attack can only be performed against the recorded or playback voice instead of real-time audio signals, thereby making them impractical for various real-world audio-domain attacking scenarios.

\textbf{Technical Preview and Contributions.} To overcome these limitations, in this paper we propose to use generative model to produce adversarial perturbations in the audio domain. This generative model learns the distribution of adversarial perturbations from training data in an offline way. Once being well-trained, the generative model can generate audio adversarial perturbations very quickly, thereby unlocking the possibility of realizing audio adversarial attack in the real-time setting. Our main contributions of this paper are summarized as follows:

\begin{itemize}

\item We, for the first time, propose a generative model-based fast audio adversarial perturbation generator (FAPG). Unlike existing methods requiring considerable adversarial perturbation generation time, our proposed FAPG generates the desired audio adversarial perturbation through a well-trained generative model Wave-U-Net \cite{stoller2018wave} in a single forward pass, thereby greatly accelerating the perturbation generation speed.

\item We propose to integrate a set of trainable class-wise embedding feature maps into FAPG to encode all the label information in the audio data to a unified model. Unlike conventional generative model-based image-domain adversarial attacks, which require different generative models for different targeted classes, the proposed audio-domain FAPG can generate adversarial perturbation targeting at any adversary-desired class using a single generator model. Such reduction significantly saves the memory cost and model training time if the adversary expects to launch attacks with multiple target classes. 


\item Built on top of the input-dependent FAPG, we further propose an input-independent universal audio adversarial perturbation generator (UAPG). UAPG is able to generate a single \textit{universal audio adversarial perturbation} (UAP), which can be applied and re-used on different benign audio inputs without the need of input-dependent perturbation re-generation. In addition, since the universality of UAP exists across different benign inputs, such important characteristic removes the prior constraint on needing to observe the entire input for perturbation generation, thereby enabling the realization of real-time audio adversarial attack.


\item We evaluate the attack performance using FAPG and UAPG against three DNN-based audio systems: speech command recognition model on the Google Speech Commands dataset~\cite{warden2018speech}, speaker recognition model on VCTK dataset~\cite{christophe2016cstr} and environmental sound classification model on UrbanSound8k dataset~\cite{Salamon:UrbanSound:ACMMM:14}. Compared with the state-of-the-art input-dependent attack, our FAPG-based attack achieves $214\times$ speedup with the comparable success rate. Compared with the existing input-independent (universal) attack, our UAPG-based attack achieves $37.22\%$ and $29.98\%$ fooling rate increase in white-box setting and black-box setting, respectively.

\end{itemize}



\section{Fast Audio Adversarial Perturbation Generator (FAPG)}

\subsection{Motivation}

\textbf{Dilemma Between Speed and Performance.}
Despite the current progress of the existing audio adversarial attacks, as analyzed in the Introduction, one of the most challenging limitations is their inherent slow generation process for adversarial perturbations. This is because: 1) the current commonly adopted underlying adversarial perturbation-generating approaches, such as PGD~\cite{madry2017towards}, C\&W~\cite{carlini2018audio} and genetic algorithms~\cite{alzantot2018did}, are built on numbers of iterations to optimize or search the perturbations. Although this iterative mechanism can bring high attack performance, the corresponding required generation time is prohibitively long, such as seconds or even hours for producing one well-crafted perturbation. 2) Reducing the number of iterations to make generation time satisfy the real-time requirement is an alternative solution; however, as shown in our experiments that will be reported later, when the iteration-based attack method is performed in a restricted time budget, the corresponding attack performance is severely degraded. 3) On the other hand, the existing one-step perturbation generation methods, e.g. FGSM~\cite{goodfellow2014explaining}, though enjoy the advantage on fast generation, suffer from the poor attack performance limitation -- they typically have much lower attack success rates than their iteration-based counterparts.

\textbf{Why Fast Perturbation Generation Matters?} 
Some readers may have questions about the necessity and motivation of the fast generation of adversarial perturbations. Why should the perturbations be generated in a real-time manner? Cannot the attacker just record the benign voice input, generate the perturbation offline under a sufficient time budget, and then play the generated adversarial audio? 
Indeed, the above hypothesized attacking strategy may fit some time-budget-relaxed scenarios; however, in practical attacking scenarios, it is more likely that the attacker does not have many opportunities to approach the victim for either recording speech or altering the victim's speech on the fly. If there is a chance, the attacker might want to record the speech, then instantly generate the adversarial perturbation (preferably using their mobile devices) and inject it onto the victim's interactive speech on the spot.
This would leave a very limited time budget and computational resource for the process of perturbation generation and injection. Thus, an efficient way to craft robust adversarial perturbations in a very timely and low computational complexity manner is highly desirable. 

\textbf{Generative Model-based Solution in Image Domain.} The above demand for fast adversarial perturbation generation is not an audio-specific problem, but also widely exists in the image domain. To satisfy this timing requirement, recent image-domain studies \cite{poursaeed2018generative,song2018constructing, phan2020cag} have proposed to utilize generative models, such as Generative adversarial network (GAN) \cite{goodfellow2014generative} and autoencoder \cite{vincent2008extracting}, to accelerate the generation of image adversarial perturbations. Different from the multi-step optimization-based approaches (e.g. C\&W and PGD), the generative model-based solutions aim to learn the distribution of adversarial perturbations from the training images. After being well-trained, the generative model performs one-step generation from input image to adversarial perturbation, where such process is essentially a fast one-pass forward propagation over the generative model, thereby significantly improving the generation speed for image adversarial perturbations.

\textbf{Challenges in Audio Domain.} Such progress on image domain naturally encourages the exploration of using generative model to accelerate audio adversarial perturbation generation. However, audio signals have a huge difference from images. A speaker's voice is essentially a 1-D time-serial signal that contains very important sequential order information. Also, unlike well-defined fixed-size image data, voice data typically have very different signal lengths even from the same user and in the same dataset. Besides these new audio-specific challenges, generative model-based audio adversarial perturbation also suffers the same class-specific model preparation problem of image-based counterparts. To be specific, when utilizing generative model to perform targeted attack, for each target class, an individual generative model has to be trained for specific use. Considering the number of classes can be very high, e.g., hundreds or even thousands, the required memory cost for launching the attack is very high.

\subsection{Proposed FAPG: Construction \& Training}
\label{subsec:FAPG}

\begin{figure}[t]
\centering
\includegraphics[width=0.45\textwidth]{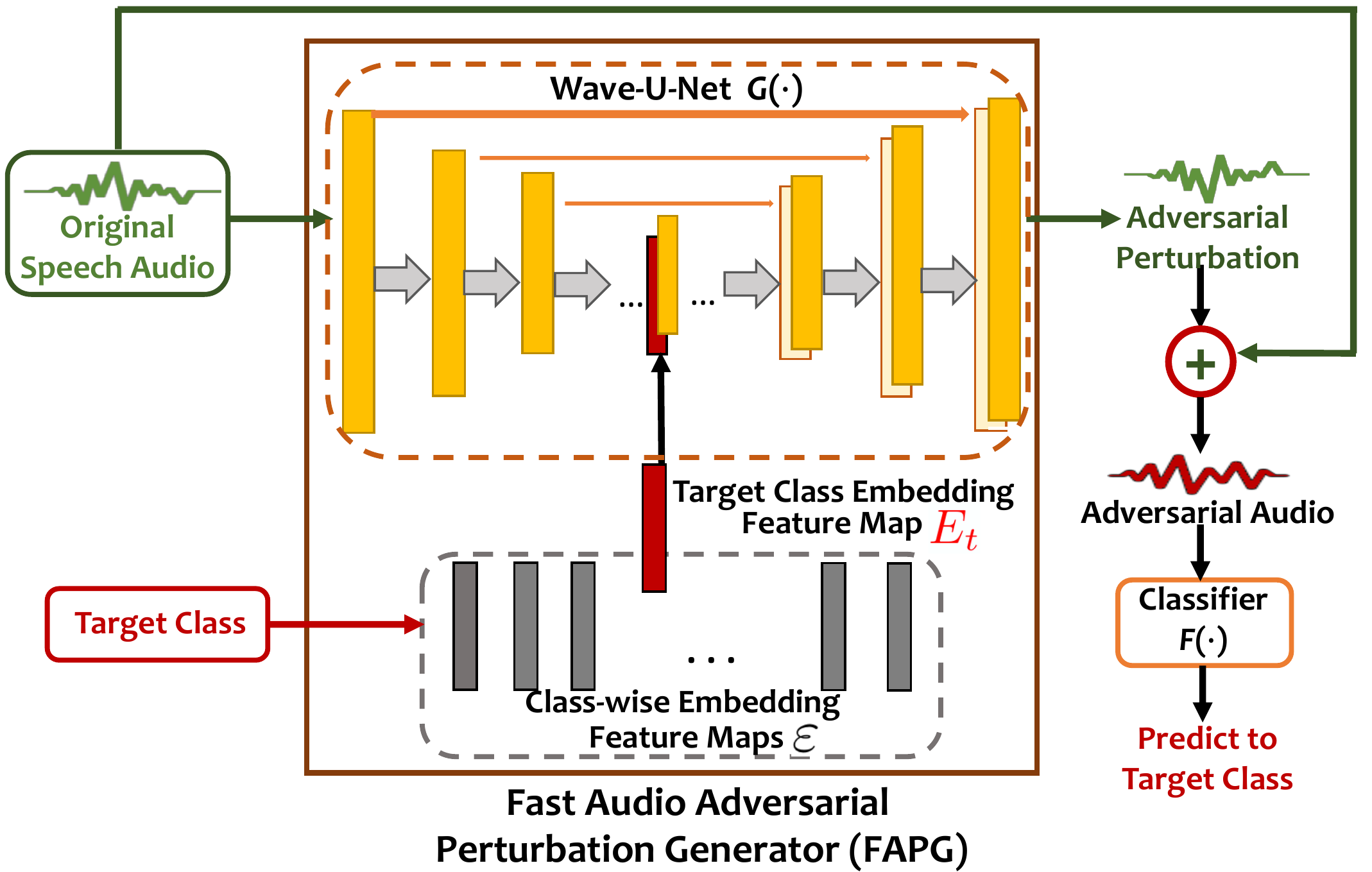}
\vspace{-3mm}
\caption{Overall architecture of the proposed FAPG.
}
\vspace{-5mm}
\label{fig:fapg}
\end{figure}

\textbf{Overall Architecture.} To address these challenges, we propose FAPG, a fast audio adversarial perturbation generator, to launch the audio-domain adversarial attack in a rapid, high-performance and low-memory-cost way. Figure \ref{fig:fapg} illustrates the overall architecture of FAPG, which contains a generative model ($G(\cdot)$), e.g., Wave-U-Net~\cite{stoller2018wave}, and multiple class-wise embedding feature maps. During the training phase, both the generative model and embedding feature maps are jointly trained on the training dataset. After proper training, given a benign audio input and a target class label $y_t$ that the adversary plans to mislead the DNN classifier ($F(\cdot)$) to, the corresponding audio adversarial perturbation can be quickly generated via performing inference of the benign input over the well-trained generative model, where the embedding feature map for the target class $y_t$ is concatenated to one intermediate feature map of $G(\cdot)$. Next, we describe the details of the used generative model and the set of embedding feature maps as follows.

\textbf{Audio-specific Generative Model.} Generative model is the core component of FAPG. Although various types of generative models have been widely used in image-domain applications, they are not well-suited for the use in FAPG due to the inherent difference between image and audio signals (e.g., sequence order and varying length). To address these challenges, we adopt Wave-U-Net \cite{stoller2018wave}, which was originally used for audio source separation, as the underlying generative model of FAPG. Wave-U-Net is a special type of CNN containing 1-D convolutional, decimal down-sampling blocks and linear interpolation up-sampling blocks. Such inherent encoder-decoder structure makes Wave-U-Net exhibit strong distribution modeling capability. Meanwhile, its unique design of first-layer 1-D convolution and up/down sampling blocks also enables Wave-U-Net can naturally capture the temporal information from 1-D varying-length data.

\textbf{Class-wise Embedding Feature Maps.} The purpose of using $k$-class embedding feature maps is to ensure that a single generative model can be re-used for attacking against different target classes instead of class-specific design. To this end, those class-aware embedding feature maps, denoted as $\varepsilon = \{E_1, E_2,...E_k\}$, are designed to be trainable, and each of them corresponds to one target class. After joint training of generative model $G(\cdot)$ and these embedding feature maps $\varepsilon$, the label information for class $y_t$ is encoded in the corresponding feature map $E_t$. Then during the generation phase $E_t$ is concatenated with one intermediate feature map of $G(\cdot)$ to craft the adversarial perturbation for the target class $y_t$. In our design, $E_t$ has the exact same shape of the intermediate feature map to which it will be concatenated. To be specific, $E_t$ is typically aligned with the intermediate feature map at the intersection between the encoder and decoder parts of Wave-U-Net. This is because the feature map has the smallest size at this position, and thereby minimizing the storage cost of the corresponding $E_t$.

\begin{algorithm}[t]
\small
\SetAlgoLined
\textbf{Require:} {Training dataset $\chi=\{x^{(1)},...,,x^{(n)}\}$, class label $\{y_1,...,,y_k\}$, DNN classifier $F(\cdot)$, noise constraint constant $\tau$} \\
\textbf{Result:} {Trained FAPG: generative model $G(\cdot)$, class-wise embedding feature maps $\varepsilon=\{E_1, ..., E_k\}$} \\
Initialize $G(\cdot)$, $\varepsilon$ and $\tau$ \\
\For{\textup{number of training iterations}}{
\For{\textup{number of steps}}{
$\boldsymbol{X} \gets$ minibatch of $m$ samples from $\chi$; \\
$y_t \gets$ $get\_random\_target \in \{y_1,...,y_k\}$; \\
$G_{t}(\cdot)\gets G(\cdot)$ embeds with $E_{t} \in \varepsilon$; \\
$\boldsymbol{\delta_t} \gets Clip(G_t(\boldsymbol{X})$, $\{-\tau, +\tau\})$; \\
$\boldsymbol{X'} \gets \boldsymbol{X} + \boldsymbol{\delta_t}$; \\
$\boldsymbol{y_{pred}}\gets F(\boldsymbol{X'})$; \\
$\text{Loss} \gets \frac{1}{m}\sum_{i}^{m} (\text{CrossEntropy}(\boldsymbol{y_{pred}^{(i)}}, y_t)+\beta \cdot\left \| \boldsymbol{\delta_t^{(i)}} \right \|_2)$; \\
  minimize {Loss} to update $G(\cdot)$ and $E_{t}$; \leavevmode\\
  decrease $\tau$\\}}  
\caption{Training Procedure of FAPG}
\label{al1}  
\end{algorithm}  
\setlength{\textfloatsep}{16pt}

\textbf{Training Procedure of FAPG.} Next we describe the training procedure of FAPG, or more specifically, the joint training for $G(\cdot)$ and $\varepsilon$. In the forward propagation phase of the entire training procedure, for each batch of input voice data $\boldsymbol{X}$, we first randomly select one target class $y_t$, and fetch the corresponding embedding feature map $E_t$. This selected feature map is concatenated into the generative model $G(\cdot)$ to form an overall model $G_t(\cdot)$. A forward pass on $G_t(\cdot)$ will be performed with input $\boldsymbol{X}$. The result, denoted as $\boldsymbol{\delta_t}$, is clipped to the range of $\{-\tau , +\tau \}$ to constrain the generated perturbation $\boldsymbol{\delta_t}$ to be imperceptible, where $\tau$ is a threshold parameter. Notice that according to our experiments, $\tau$ should be set as a relatively large value initially, and gradually decreased during the training procedure. Empirically such adjusting scheme can bring better training convergence.

After perturbation $\boldsymbol{\delta_t}$ is calculated from the generative model, it is imposed on the benign data to form the adversarial input, which can cause the misclassification of DNN classifier $F(\cdot)$. Then, the loss function, which is the key of the entire training procedure, is formulated as follows:

\begin{equation}
\setlength{\abovedisplayskip}{3pt}
\setlength{\belowdisplayskip}{3pt}
\begin{aligned}
    Loss(\boldsymbol{X}, y_t) = -y_t \cdot log(F(\boldsymbol{X}+ G_t(\boldsymbol{X})))+ \beta \cdot \|G_t(\boldsymbol{X})\|_2,
\end{aligned}    
\end{equation}
where the first and second terms are the cross-entropy loss and $L_2$ loss, respectively, and $\beta$ is a pre-set coefficient. The existence of $L_2$ loss in the entire loss function is to control the attack strength and make the generated adversarial perturbation imperceptible.

Consequently, in the backward propagation phase both the generative model $G(\cdot)$ and the current selected embedding feature map $E_{t}$ are updated simultaneously by minimizing the loss function. Notice that for each batch of data, $E_{t}$ is randomly selected. Therefore after rounds of iterations the generative model $G(\cdot)$ itself learns the general distribution of adversarial perturbations, and different $E_{t}$ learns the encoded information for each specific target class. The details of the entire FAPG training procedure are summarized in Algorithm \ref{al1}.

\section{Universal Audio Adversarial Perturbation Generator (UAPG)}
\label{sec:universal}
\vspace{-1mm}



\begin{figure}[t]
\centering
\includegraphics[width=0.4\textwidth]{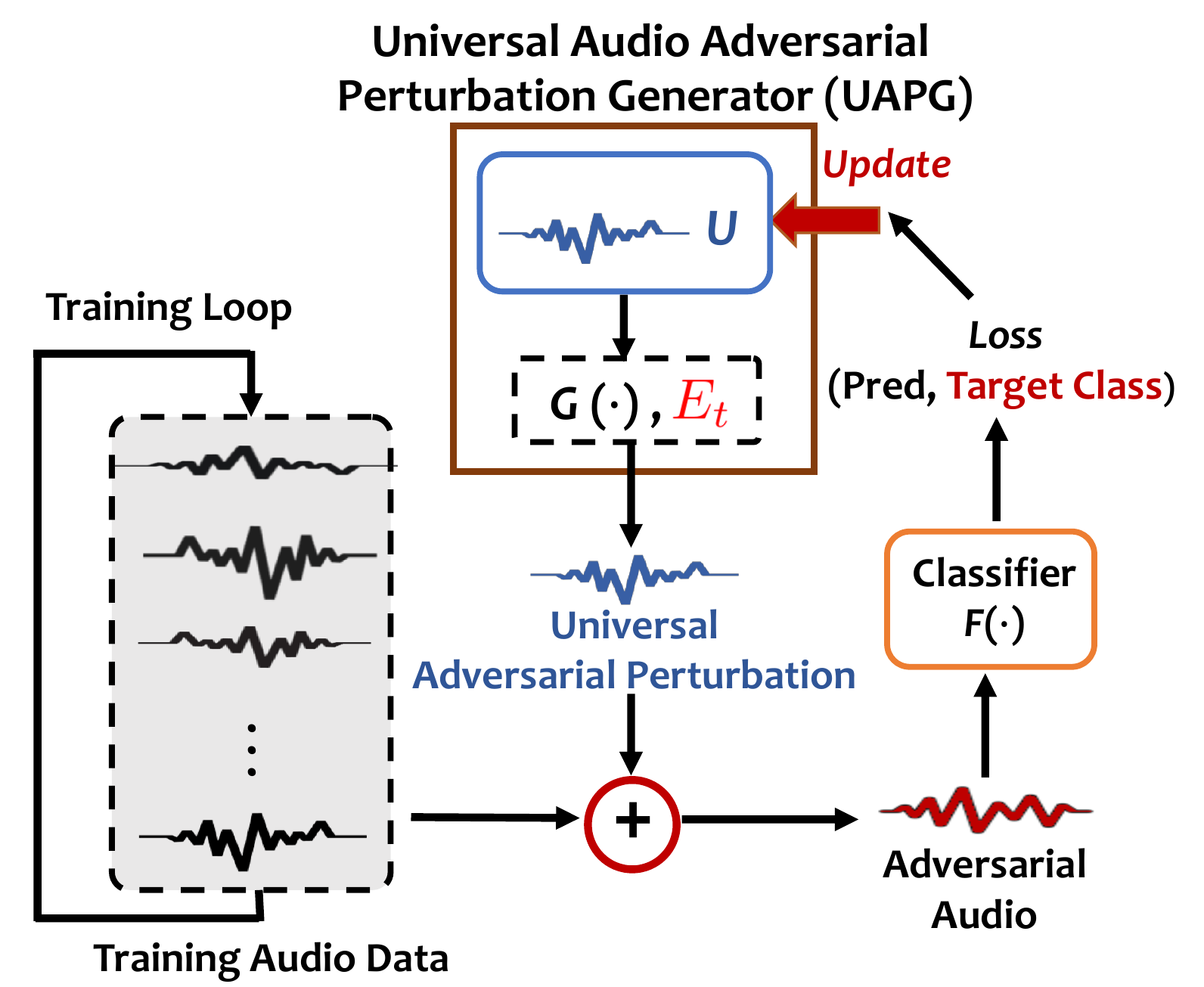}
\vspace{-3mm}
\caption{Overall architecture and training scheme of the proposed UAPG.
}
\vspace{-5mm}
\label{fig:uapg}
\end{figure}

\subsection{Motivation}
\textbf{Reducing the Observation of Full Content -- Why It Matters?} As presented in the previous section, FAPG provides a fast solution to generate audio adversarial perturbations. However, it is essentially an input-dependent generating approach. In fact, most of the state-of-the-art adversarial attack methods, in both audio and image domains, belong to the input-dependent attack category. In other words, the underlying perturbation generation mechanism is based on the observation of the entire benign input. 
Although such assumption may hold for most image processing applications, it is very challenging to satisfy such requirement in the practical real-time audio applications. This is because audio signals have inherent temporal sequence, it is unrealistic to pre-know the full content of the ongoing voice input during the input-streaming phase. In other words, the attacks can only be performed against the recorded or playback voice, thereby severely limiting the attacking feasibility and scenarios. Consequently, besides significantly reducing the perturbation generation time, a practical audio adversarial attack should also reduce the demand of observing the content of benign input as low as possible. 

\textbf{Universal Audio Adversarial Perturbation Generator (UAPG).} To achieve this, we further develop universal audio adversarial perturbation generator (UAPG) to craft audio-domain universal adversarial perturbation (UAP). As revealed by its name, a single universal adversarial perturbation can be applied and re-used on different benign inputs to cause mis-classification without the need of input-dependent perturbation re-generation. Such unique universality completely removes the prior constraint on observation of the entire input, and makes UAPG very suitable to launch real-time audio adversarial attack with zero time cost.

\textbf{Challenges of UAPG Design.} The attractive benefits of UAP have already led to some efforts on studying image-specific UAP~\cite{moosavi2017universal,poursaeed2018generative}. Lending the methodology used in those research progress in image domain, recent works~\cite{vadillo2019universal,neekhara2019universal} report audio-domain UAP generating methods for speech command recognition and speech-to-text systems, respectively. Besides, \cite{DBLP:journals/corr/abs-1905-13399} also proposes a technique to realize real-time audio adversarial attack without using the entire voice input, which has the similar effect as using UAP.

Despite these existing efforts, designing a robust and powerful UAPG is still non-trivial but facing two main challenges: 1) the experimental results show that the current audio-domain UAPs typically have much lower attack performance than the input-dependent perturbations; and 2) the attack enabled by some audio-domain UAPs are only untargeted attack, where the adversary cannot precisely obtain the desired target results.

\subsection{Proposed UAPG: Construction \& Training}
\label{subsec:UAPG}

\textbf{Overall Scheme.} Different from the existing studies, we aim to devise a UAPG which can achieve high targeted attack performance. Figure 2 shows the  key idea: we produce an input-dependent UAP based on a signal vector $U$, which is to be trained to exhibit a certain degree of universality. After initialization, $U$ is used to produce UAPs, and it is updated in an iterative way by gradually improving the universality of the derived UAPs across different training data samples. Finally, an effective UAPG is able to be constructed via evolving well-trained $U$.

\textbf{From FAPG to UAPG.} 
The underlying method used for generating UAPs is our proposed FAPG. Intuitively, FAPG learns to estimate the distribution of adversarial perturbations instead of iteratively optimizing the perturbation for a specific audio input. Therefore the FAPG-generated
perturbation naturally exhibits better universality than that the one comes from the non-generative method. Moreover, our FAPG is designed to integrate various target classes information into a single generative model, thereby enabling the capability of producing targeted universal perturbations. 

\textbf{Training Procedure of UAPG.} 
We then introduce the training details to facilitate an effective UAPG. In general, to formulate an input-agnostic universal attack, our goal is to find a universal perturbation $\upsilon_t$ to satisfy: 
\begin{equation}
argmax \; F(\boldsymbol{x^{(i)}}+ {\upsilon_t})=y_t \textup{ for most}\; x \;\sim \; \chi.
\end{equation}
The training procedure of UAPG is shown in Algorithm \ref{al2}. We aim to generate a single universal perturbation $\upsilon_t$ via the well-trained $G(\cdot)$ and the corresponding $E_t \in \varepsilon$, which can be obtained from the well-trained input-dependent FAPG. Different from input-dependent scenario, the audio input signal is now replaced by a single trainable vector $U$. Then the universal perturbation is returned and imposed on the benign data to craft the adversarial audio example. Through feeding such an adversarial audio into the DNN classifier $F$, we can update $U$ by minimizing the following loss function:
\begin{equation}
Loss =  -y_t \cdot F(\boldsymbol{X}+ G_t(U)) + \beta \cdot   \|G_t(U)\|_2,
\label{equ:3}
\end{equation}
where the first and second terms represent the cross-entropy loss and $L_2$ loss, respectively. With the guidance of the above loss function, we optimize $U$ by iteratively applying the derived $\upsilon_t$ across the entire training data. In particular, in order to construct a UAPG that can be universally applied to any target class, at each training step, a random target class is selected to help $U$ to learn inter-class representations. After constructing the unified $U$, the universal perturbations computed by our UAPG can be effectively applied on any input data to fool the DNN model in an audio-agnostic way, without requiring re-generating adversarial perturbation for each individual audio input.

\begin{algorithm}[t]
\small
\caption{Training Procedure of UAPG}
\label{al2}
\SetAlgoLined
\textbf{Require:} {Training dataset $\chi=\{x^{(1)},...,,x^{(n)}\}$, class label $\{y_1,...,,y_k\}$, DNN classifier $F(\cdot)$, generative model $G(\cdot)$, class-wise embedding feature maps $\varepsilon$, noise constraint constant $\tau$ }\\
\textbf{Result:} Trained \textup{UAPG}\\
Random initialize $U$\\
\For{\textup{number of training iterations}}{
\For{\textup{number of steps}}{
  $y_t \gets$ $get\_random\_target \in \{y_1,...,y_k\}$;\\
  $G_{t}(\cdot)\gets G(\cdot)$ embeds with $E_{t} \in \varepsilon$ ;\\
  $\textup{UAP}\; {\upsilon_t} \gets Clip(G_t(U)$, $\{-\tau, +\tau\})$ ;\\
  $\boldsymbol{X} \gets$ minibatch of $m$ samples from $\chi$;\\
  $\boldsymbol{X'} \gets \boldsymbol{X} + \upsilon_t$ ;\\
  $\boldsymbol{y_{pred}}\gets F(\boldsymbol{X'})$ ;\\
    $\text{Loss} \gets \frac{1}{m}\sum_{i}^{m} (\text{CrossEntropy}(\boldsymbol{y_{pred}^{(i)}}, y_t)+\beta \cdot\left \| \boldsymbol{\upsilon_t} \right \|_2$);\\
  minimize {Loss} to update $U$;\\
  }} 
\end{algorithm}

\section{Attack Evaluation}
\label{sec:exp}
\subsection{Experimental Methodology}

\subsubsection{Target Model and Dataset.}
We evaluate the proposed FAPG and UAPG on three types of the DNN-based audio systems, namely, \textit{speech command recognition}, \textit{speaker recognition}, and \textit{environmental sound classification}.
\begin{itemize}
\item \textbf{Speech Command recognition.} We use a convolutional neural network (CNN)-based speech command recognition model (CNN-trad-fpool3) as proposed in~\cite{sainath2015convolutional}, which has served as the target model in many previous studies~\cite{alzantot2018did,abdoli2019universal,yu2018towards}. The network is trained on a crowd-sourced speech command dataset~\cite{warden2018speech}, which contains $46,278$ utterances from 10 representative speech commands 
sampled at $16kHz$, with each recording being cropped to $1$s. $40$-dimensional MFCC features are extracted as the input of the model. We randomly separate the dataset into training and testing set with a ratio of $4$ to $1$, and the recognition accuracy of this baseline model on the testing dataset is $89.2\%$.

\item \textbf{Speaker Recognition.} A pre-trained X-vector model\footnote{Avaible at \url{https://kaldi-asr.org/models/m3}}~\cite{snyder2018x} with DNN-based embedding model and probabilistic linear discriminant analysis (PLDA) backend is used as the target speaker recognition model. The features are $30$-dimensional MFCC features with a frame length of $25ms$. The dataset we use is an English multi-speaker corpus provided in CSTR voice cloning toolkit (VCTK)~\cite{christophe2016cstr}, which contains $44217$ utterances spoken by $109$ speakers, with each recording being cropped to $1.75$s. The speakers are enrolled utilizing $80\%$ of the data, while the rest is reserved for testing. This results in a baseline accuracy of $92.8\%$ on $8896$ testing utterances from $109$ speakers.

\item \textbf{Environmental Sound Classification.} A $1$-dimensional CNN model (referred as \textit{CNNrand} in \cite{abdoli2019end}) is used as the target model. The model is trained on the UrbanSound8k dataset~\cite{Salamon:UrbanSound:ACMMM:14}, which contains a total number of $8732$ audio clips from $10$ different environmental scenes. Each recording is cropped to $50999$ samples which corresponds to roughly $3$ seconds at $16$ kHz. The dataset is split into training, validation and testing set with a ratio of $8:1:1$. After training, the classification accuracy on $10$ classes is $83.4\%$.
\end{itemize}

\subsubsection{Evaluation Metrics.} (1) \textit{Fooling Rate (FR)} is used for evaluating both targeted and untargeted attacks, which shows the ratio of the number of adversarial examples that lead to a false classification and the total number of adversarial examples; (2) \textit{Success Rate (SR)} is only used for evaluating targeted attacks, which is the ratio of the number of attacks resulting in the adversarial example being classified as the target class and total attack attempts; (3) \textit{Distortion Metric:} We quantify the relative noise level of $\delta_t$ with respect to the original audio $x_i$ in decibels (dB): $D(x_i, {\delta}_t) = 20log_{10} \frac{max({\delta}_t)}{max(x_i)}$.





\subsection{Audio-dependent Targeted Attack via FAPG}

\textbf{FAPG Generator Implementation.} We use model $M1$ of Wave-U-Net~\cite{stoller2018wave} to construct our FAPG. Specifically, our model contains $5$ down-sampling blocks and $5$ up-sampling blocks. The feature map size of the last encoding layer is also the size of each additional class-wise embedding feature map.
For FAPG, a total of 10,000 training steps are conducted using Adam optimizer with the batch size of 100. The initial learning rate is set to $1e^{-4}$ and gradually decayed to $1e^{-6}$. $\beta$ is set as 0.1 for all dataset. $\tau$ is initially set as 0.1 and reduces to 0.05 and 0.03 at step of 3,000 and 7,000 for command recognition and speaker recognition, and it stops reducing as 0.05 for sound classification model, which leads to an approximate noise level of $-30$ dB and $-18$ dB respectively.

\begin{table}[]
\centering
\resizebox{\linewidth}{!}
{\begin{tabular}{|c|c|c|c|c|}
\hline
                                                       &      FGSM &PGD     & C\&W    & FAPG      \\ \hline
\begin{tabular}[c]{@{}c@{}} Command\\Recognition \end{tabular}    & 11.89\% & 11.96\% & 13.26\% & \textbf{97.77\%}   \\ \hline
\begin{tabular}[c]{@{}c@{}}Speaker \\ Recognition\end{tabular}& 1.65\% & 0.96\% & 11.09\% & \textbf{98.35\%}   \\ \hline
\begin{tabular}[c]{@{}c@{}}Sound \\ Classification\end{tabular}&14.46\% &   10.08\%      & 11.42\% &  \textbf{92.93\%} \\ \hline
\end{tabular}}
\caption{Success rate (SR) of audio-dependent targeted attacks under constrained time budget ($0.065$s).}
\label{tab:fapg_1}
\end{table}


\textbf{Attack Speedup and Performance.}
To demonstrate the ability of the proposed FAPG in terms of achieving high success rate while maintaining a short attack generation time, we conduct experiments on the three aforementioned target models under different time conditions.
Table~\ref{tab:fapg_1} compares the attack performance of the proposed FAPG with commonly-used attacks, i.e., FGSM~\cite{goodfellow2014explaining}, PGD~\cite{madry2017towards} and C\&W~\cite{carlini2017towards} under constrained time budget scenario, which requires to generate adversarial example with no more than $0.065$s (the approximate execution time for one iteration in PGD and C\&W attack. For fair comparison, we constrain the perturbations generated by these attacks with an infinity norm of $0.03$ for speech command classification and speaker recognition, and $0.05$ for environmental sound classification, which are the same as used in FAPG implementation. As shown in Table~\ref{tab:fapg_1}, the proposed FAPG can achieve high attack success rate (SR) (over  $90\%$) under the short time budget for all the three target models, while FGSM, PGD and C\&W attack can only achieve less than $15\%$ SR with limited attack time budget.

Additionally, we also conduct experiments when sufficient time budget is granted. As shown in Table~\ref{tab:fapg_2}, though PGD and C\&W achieve the very similar SRs to our proposed FAPG, they require much longer adversarial perturbation generation time. For instance, for speaker recognition task PGD needs $4.33$s and C\&W even requires more than $10$s to launch the attack, while the time period for each data is only $1.75$s. Such huge gap makes the PGD and C\&W-based attacks infeasible in the practical real-time attack scenarios. On the other hand, with achieving the very similar high SR, our proposed FAPG only needs $0.05$s to generate adversarial perturbation, thereby bringing very high speedup (up to $86\times$ and $214\times$ as compared with PGD and C\&W, respectively). Also, compared with another fast generation approach FGSM, FAPG achieves much higher SR.

\textbf{Memory Cost Reduction.} Our proposed trainable class-wise feature maps can reduce the memory cost significantly. Without the class-wise feature embedding maps, launching targeted attack requires to train one generative model for each target class, which results in a memory consumption of $23.8$ MB, $259$ MB, and $23.8$ MB for the speech command recognition, speaker recognition, and sound classification model, respectively. In contrast, by utilizing the class-wise embedding feature maps, our proposed FAPG only requires to train a single generative model and a set of embedding maps, regardless of the number of target classes, and therefore only takes up $2.4$ MB, $3.53$ MB, and $2.44$ MB for these three target models respectively. This leads to a memory cost reduction of $9.9\times$, $73.5\times$, and $9.8\times$, respecitively.

\begin{table}[]
\centering
\resizebox{1\linewidth}{!}
{\begin{tabular}{|p{1.75cm}|p{0.8cm}|c|c|c|c|}
\hline
                                                            &  Metric     &    FGSM          & PGD     & C\&W    & FAPG      \\ \hline
                                                                   
\multirow{2}{*}{\begin{tabular}[c]{@{}c@{}}Command\\Recognition \end{tabular}} & SR(\%)     &  11.89 & 96.03 & \textbf{97.92} & 97.77 \\ \cline{2-6} 
                                                          &  Time 
&    0.05s    & 1.36s   & 9.16s   & \textbf{0.05s}    \\ \hline
\multirow{2}{*}{\begin{tabular}[c]{@{}c@{}}Speaker \\ Recognition\end{tabular}} &SR(\%) & 1.65 &    97.47 &  98.08     & \textbf{98.35}  \\ \cline{2-6} 
        &Time  
        &      0.05s   &  4.33s      &    10.74s    & \textbf{0.05s}     \\ \hline
\multirow{2}{*}{\begin{tabular}[c]{@{}c@{}} Sound \\ Classification\end{tabular}}&SR(\%) & 14.46 &      91.74   & 92.55 & \textbf{92.93} \\ \cline{2-6} 
                                          & Time 
                                          &   0.05s   &  1.85s       &   4.69s       & \textbf{0.05s}     \\ \hline
\end{tabular}}
\caption{Success rate (SR) and the corresponding attack generation time of audio-dependent targeted attacks under sufficient time budget. }
\label{tab:fapg_2}
\end{table}

\begin{table*}
\centering
\resizebox{0.85\linewidth}{!}{
\begin{tabular}{|c|c|c|c|c|c|c|}
\hline
      & \multicolumn{2}{c|}{Command Recognition }                 & \multicolumn{2}{c|}{Speaker Recognition}        & \multicolumn{2}{c|}{Sound Classification}   \\ \hline
Attack Method  & \begin{tabular}[c]{@{}c@{}}UAP-HC\end{tabular} & UAPG             & \begin{tabular}[c]{@{}c@{}}RURA\end{tabular} & UAPG             & \begin{tabular}[c]{@{}c@{}}UAAP\end{tabular} & UAPG      \\ \hline
 FR & 52.78\%                                                                    & \textbf{90.03\%} & N/A                                                       & \textbf{90.05\%} & N/A                                                          & \textbf{91.01\%} \\ \hline
 SR & N/A                                                             & \textbf{89.90\%} & 86.17\%                                                              & \textbf{89.59\%} & 85.40\%                                                                  & \textbf{86.05\%} \\ \hline
\end{tabular}
}
\caption{Success rate (SR) of audio-agnostic targeted attacks under white-box setting.}
\label{tab:universal}
\end{table*}

\subsection{Audio-agnostic Universal Attack via UAPG}
\textbf{UAPG Implementation.} 
The proposed UAPG is built on a pre-trained FAPG model and a trainable universal adversarial input vector $U$ with the same size of the original audio input. The vector $U$ is then trained on the same training set as used for the target model training. A total number of $8000$ training steps are conducted using Adam optimizer with a learning rate of $1e^{-4}$ and a batch size of $100$. We set $\tau$ to $0.03$ which corresponds to an average distortion of $-30.21dB$ of the generated adversarial perturbations for speech command recognition model and speaker recognition, and $\tau=0.05$ for environmental sound classification. 

\begin{figure}[t]
\centering
\includegraphics[width=0.45\textwidth]{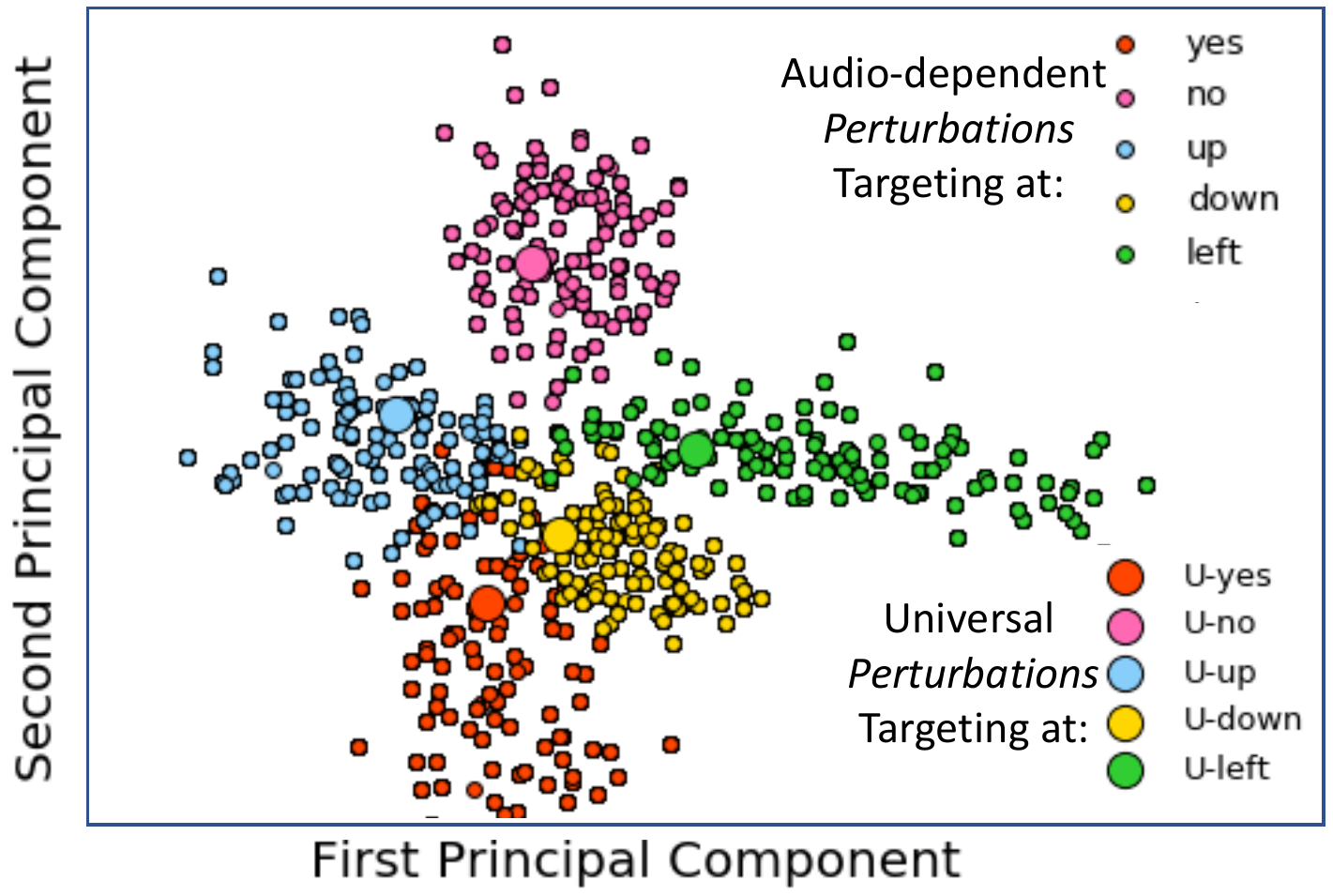}
\caption{Visualization of audio-dependent perturbations and universal perturbations targeting at different speech commands.}
\label{fig:pca}
\end{figure}

\textbf{Analysis of Learned Representation.}
To investigate the effectiveness of UAPG, we plot the audio-dependent perturbations generated by FAPG as well as the audio-agnostic perturbations generated by UAPG on the speech command recognition model using principal component analysis (PCA)~\cite{wold1987principal}. We show the adversarial perturbations targeting at five commands in Figure~\ref{fig:pca}. Although the universal perturbations are created without accessing the distribution of real speech commands, all universal perturbations locate within the manifold of corresponding audio-dependent perturbations generated for the same target class. This demonstrates that our UAPG can efficiently learn the inherent adversarial representations with respect to each target command.  

\textbf{White-box Attack Performance.} We compare the performance of our proposed UAPG with several state-of-the-art audio universal attacks, including UAP-HC~\cite{vadillo2019universal} which is based on DeepFool algorithm~\cite{moosavi2016deepfool}, RURA~\cite{xie2020real}, and UAAP~\cite{abdoli2019universal}. To evaluate UAPG attack, for each target model, we generate one universal perturbation for each target class.
Table \ref{tab:universal} presents the results of UAP-HC on the speech command model, RURA on the speaker recognition model, UAAP on the sound classification model, and the proposed UAPG on all three of these models. Specifically, since UAP-HC is designed to be an untargeted universal attack, we only report its average fooling rate (FR). 
We observe that our proposed UAPG outperforms existing methods on all three tasks, with an average SR of $89.90\%$, $89.59$, and $86.05\%$ when evaluated on the three models respectively.

\textbf{Black-box Attack Performance.} We also evaluate the performance of the proposed UAPG under black-box settings, where the architecture and parameters of the target victim model is unknown. For each task, we first train UAPG on a substitute model (CNN-3 model~\cite{zhang2017hello}, d-Vector~\cite
{variani2014deep}, EnvNetV2~\cite{tokozume2017learning}) as is shown in Table~\ref{tab:transfer}, and then evaluate the generated adversarial examples on the target model to study its transferbility.
For the speech command recognition model, we compare the performance of the proposed UAPG with the recent untargeted real-time adversarial attack (RAA)~\cite{DBLP:journals/corr/abs-1905-13399} on the same target model in a black-box manner. As shown in Table~\ref{tab:transfer}, our proposed UAPG achieves high FR even when tested in a black-box setting on different tasks. Compared with the state-of-the-art untargeted real-time attack RAA, our UAPG achieves $29.98\%$ FR increase.

\begin{table}[]
\resizebox{\linewidth}{!}{
\begin{tabular}{|c|c|c|c|c|}
\hline
                                                            & \multicolumn{2}{c|}{\begin{tabular}[c]{@{}c@{}}Speech\\ Recognition\end{tabular}} & \begin{tabular}[c]{@{}c@{}}Speaker\\ Recognition\end{tabular} & \begin{tabular}[c]{@{}c@{}}Sound\\ Classification\end{tabular} \\ \hline
\begin{tabular}[c]{@{}c@{}}Substitude \\ Model\end{tabular} & \multicolumn{2}{c|}{CNN-3}                                                        & d-Vector                                                      & EnvNetV2                                                       \\ \hline
\begin{tabular}[c]{@{}c@{}}Target\\ Model\end{tabular}      & \multicolumn{2}{c|}{CNN-trad-fpool3}                                                             & X-Vector                                                      & CNNrand                                                        \\ \hline
Method                                                      & \begin{tabular}[c]{@{}c@{}}RAA  \end{tabular}                            & UAPG                                    & UAPG                                                          & UAPG                                                           \\ \hline
FR                                                          & 43.5\%                                  & \textbf{73.48\%}                                 & 80.50\%                                                       & 69.26\%                                                      \\ \hline
\end{tabular}}
\caption{Fooling rate (FR) of audio-agnostic targeted attacks under black-box setting. RAA~\cite{DBLP:journals/corr/abs-1905-13399} only reports result on speech command recognition task.}
\label{tab:transfer}
\end{table}

\section{Conclusion}

In this work, we propose a fast and universal adversarial attack on three audio processing systems: speech command recognition, speaker recognition and environmental sound classification. By exploiting Wave-U-Net and the class-wise feature embedding maps, our proposed FAPG can launch fast audio adversarial attack targeting at any speech command using a unified generative model within a single pass of feed-forward propagation, which results in an adversarial perturbation generation speedup up to $214\times$ comparing to the state-of-the-art solutions. Moreover, built on the top of FAPG, our proposed UAPG is able to generate universal adversarial perturbation that can be applied on arbitrary benign audio input. Extensive experiments demonstrate the effectiveness of the proposed FAPG and UAPG.

\clearpage
\section{Acknowledgments}
This work is partially supported by Air Force Research Lab (AFRL) under Grant No. FA87501820058,  the Army Research Office (ARO) grant W911NF1910405 and National Science Foundation (NSF) award CCF-1937403, CCF-1909963, CCF-2028876 and CNS-1820624.

\bibliography{aaai21}

\end{document}